\documentstyle[12pt]{article}
\begin{document}
\font\bbb = msbm10
\def\Bbb#1{\hbox{\bbb #1}}
\font \lbb = msbm7
\def\LBbb#1{\hbox{\lbbb #1}}

\newcommand{\BB} {{\Bbb B}}
\newcommand{\CC}   {{\Bbb C}}
\newcommand{\RR}  {{\Bbb R}}
\newcommand{\ZZ} {{\Bbb Z}}
\newcommand{\EE} {{\Bbb E}}
\newcommand{\NN} {{\Bbb N}}
\baselineskip 22pt plus 2pt
\begin{center}
{\bf Empirical State Determination of Entangled Two-Level Systems and
its Relation to Information Theory\\ \ \\
Y.\ Ben-Aryeh and A.\ Mann}\\
Department of Physics, Technion--Israel Institute of Technology,\\
Haifa 32000, Israel\\[0.25cm]
{\bf B.\ C.\ Sanders}\\
Department of Physics,
Macquarie University,
Sydney,
New South Wales 2109, Australia\\ \ \\
\end{center}

\noindent{\bf Abstract}

Theoretical methods for empirical state determination of entangled two-level
systems are analyzed in relation to information theory. We show that hidden 
variable theories would lead to a Shannon index of correlation between the 
entangled subsystems which is larger than that predicted by quantum 
mechanics. Canonical representations which have maximal correlations are 
treated by the use of Schmidt decomposition of the entangled states, 
including especially the Bohm singlet state and the GHZ entangled states. 
We show that quantum mechanics does not violate locality, but does 
violate realism.
\pagebreak

\noindent{\bf 1. Introduction}

In the present paper we would like to treat the problem of empirical 
determination of the quantum state of entangled two-level systems. We limit 
here the discussion to pure quantum states and use the criterion 
$\rho^2=\rho$ for pure states, where $\rho$ is the density matrix of the 
system.  The problem of empirical determination of quantum states was 
treated by Band and Park [1,2]. They treated various systems and 
various kinds of measurements. Since their formalism is quite general it 
turns out to be quite complicated. Recently there have been important 
developments in relation to measurements of the quantum state of light [3]. 
As we treat in the present article the special case of entangled two-level 
systems and the special case of measurements related to Pauli spin operators, 
we can adopt here a different, simpler formalism.  Practical 
measurements of two-level systems depend on the specific system and can be 
made for example by Stern-Gerlach devices for spin-$\frac{1}{2}$ systems 
[4], dipole moments and inversion of population measurements for two-level 
atoms [5], polarization states for entangled photons [6], etc. Our general 
approach to measurement of entangled two-level systems is 
similar to others [1-3] which have treated different systems:
\begin{quote}
``In our discussion, the term quantum state refers to the ensemble not to 
any individual system (An ensemble may be generated, however, by taking a 
single system alternately prepared, measured, identically prepared, etc.); 
the state determination is synonymous with the determination of the density 
$\rho$'' [2].
\end{quote}

We follow here the approach that the state 
function (or more generally the density matrix) represents our state of 
knowledge of the system [4]: ``Once it is accepted it is not surprising 
that new information can change the state of our knowledge. It is also 
clearly no difficulty that a measurement performed in one region of space 
can give us information about an object which is far away without this 
implying the transmission of influence instantaneously [4]''. We assume the 
``epistemological'' character of the wavefunction in the sense that it is a 
device for making statistical predictions for future experiments on the 
basis of our present knowledge of the system. (Interesting studies which 
give ``ontological'' meaning to the wavefunction, i.e., it exists as a real 
physical wave independent of our knowledge of it, were also developed 
[7-10]). 

Recently, with new developments of experimental methods, a number of 
possible practical applications of quantum entangled states have been 
proposed, including
quantum computation [11,12] and quantum teleportation [13]. Entangled states
with two particles have been employed to test Bell's inequality and to rule out
local-realistic descriptions of nature [14]. Entangled states with three 
particles, the so-called Greenberger-Horne-Zeilinger (GHZ) states [15], and with
more particles [16,17] have been proposed for studying the role of quantum 
correlations, and there is a large current interest in carrying out experiments
on such states. It is relatively easy to produce entangled states for
photons, e.g., by parametric amplifiers [6], but entanglement of two atoms
(massive particles) has been produced for the first time only in 1997 [18].
Various processes for the production of entangled states have been studied 
including, among others, the production of entangled coherent states [19].
The phenomenon of entanglement is essentially related to the problem of
empirical determination of the entangled quantum states which is the main
topic of the present article.

The paper is arranged as follows: In Sec.~2 we show that the assumption of
hidden variables for the correlation of entangled states gives an amount of
information which is incompatible with certain results obtained by QM [20].
We use here the concept of relative state introduced by Everett [21,22], and
follow some of his mathematical derivations, but do not support his 
physical conclusions. Our main approach to empirical determination of 
entangled two-level states is analyzed in Sec.~3. We show in this section 
that there is no ``quantum nonlocality'' problem, at least not for the 
entangled two-level systems. It was already pointed out [23] ``that the 
idea of nonlocal influencing of one particle on another when they are in 
space-like separated regions has neither empirical nor theoretical 
support''. We support this viewpoint from another perspective, and show by 
our analysis that QM does not violate locality but violates realism even 
for single particles.
\pagebreak

\noindent{\bf 2. On the Relation between the Shannon Index of Correlation
and Hidden Variable Theories for Two-Component Systems}

In the articles of Everett on ``The Many-Worlds interpretation of quantum mechanics''
[21,22], the EPR problem has been related to the concept of ``relative state
function''. In this description one considers a composite system
$S=S_1 + S_2$, in the state $\psi^S$. To every state $\eta$
of $S_2$ a state $\psi^\eta_{\rm rel}$ is associated which is called the
relative state in $S_1$ for $\eta$ in $S_2$, through
\begin{eqnarray}
\psi^\eta_{\rm rel} = N \sum_i \langle \phi_i\eta | \psi^S\rangle | \phi_i\rangle \ , 
\end{eqnarray}
where $\{\phi_i\}$ is any complete orthonormal set in $S_1$ and $N$ is a
normalization constant. An important property of $\psi^\eta_{\rm rel}$
is its uniqueness, i.e., it is independent of the choice of the basis
$\{\phi_i\}$ [21,22]. Another important property of the relative state is
that $\psi_{\rm rel}^\eta$ gives the conditional expectations of all operators,
conditioned by the state $\eta$ in $S_2$ [21,22]. 
Everett concludes by following
his analysis: ``It is meaningless to ask the absolute state of a subsystem --
one can only ask the state relative to a given state of the remainder 
system''. By following such analysis one enters into the problem of quantum 
correlations between two separated subsystems.

The canonical correlation, which describes the fundamental correlation between
two separated subsystems $S_1$ and $S_2$, is obtained by choosing a 
representation in which both reduced density matrices $\rho^{S_1}$ and 
$\rho^{S_2}$ of the subsystems are diagonal. In this representation the 
state $\psi^S$ is described by the Schmidt decomposition [21,22,24]:
\begin{eqnarray}
\psi^S=\sum_i a_i \zeta_i\eta_i \ , 
\end{eqnarray}
where the $\{\zeta_i\}$ and $\{\eta_i\}$ constitute orthonormal sets
of states for $S_1$ and $S_2$, respectively. Any pair of operators 
$\hat{A}$ in $S_1$ and $\hat{B}$ in $S_2$, which have as non-degenerate
eigenfunctions the set  $\{\zeta_i\}$ and $\{\eta_j\}$ (i.e., operators
which define the canonical representation [21,22]) are ``perfectly'' 
correlated in the sense that there is a one-one correspondence between 
their eigenvalues. The probability for eigenvalues $\lambda_i$ of 
$\hat{A}$ and $\mu_j$ of $\hat{B}$ is given by
\begin{eqnarray}
P(\lambda_i \ {\rm and} \ \mu_j) = P_{ij} 
\end{eqnarray}
The Shannon index of correlation in this representation is given by [20]:
\begin{eqnarray}
I_{\rm Shann} = \sum_{i,j} P_{ij} \log \left(\frac{P_{ij}}{P_iP_j}\right) 
\end{eqnarray}
Classically, one can consider two random variables $X$ and $Y$ with a joint
probability $p(x,y)$ and marginal probabilities $p(x)$ and $p(y)$.
Then the mutual information $I(X,Y)$ is given by [25,26]:
\begin{eqnarray}
I(X,Y) = \sum_{x,y} p(x,y)\log\left[\frac{p(x,y)}{p(x)p(y)}\right] 
\end{eqnarray}
Quantum mechanically one considers the Shannon index of correlation [20] which is 
the analog of Eq.~(5) but which depends on the representation of the 
quantum state. The canonical representation (Eq.~(2)) gives the 
maximal Shannon index of correlation (Eq.~(4)) [20-22].

One should distinguish between the Shannon index of correlation, and the quantum
index of correlation -- which is related to the quantum entropy. The overall state
of the two-component system is described by a density operator $\rho$ and the 
states of the component systems are described by the reduced density operators
$\rho_a$ and $\rho_b$. Using the definition of entropy
\begin{eqnarray}
S=-{\rm Tr} \rho\ln\rho \ , 
\end{eqnarray}
the quantum index of correlation for a two-component system is [20]:
\begin{eqnarray}
I_c=S_a+S_b-S 
\end{eqnarray}
where
\begin{eqnarray}
S_a = - {\rm Tr}\rho_a\ln\rho_a \ : \ \ S_b=-{\rm Tr} \rho_b\ln\rho_b 
\end{eqnarray}
As we restrict the discussion to pure states,
the entropy $S$ is precisely
zero. Barnett and Phoenix [20] derived the important result that, for
a pure two-component system the observation of the Shannon index of 
correlation, for a certain representation, cannot provide more information 
than half the information contained in the quantum index of correlation. As 
one can easily verify, the canonical representation given by Eq.~(2) gives 
the maximal Shannon index of correlation $I_{\rm Shann} = \frac{1}{2} I_c = 
\frac{1}{2}(S_a + S_b)$, where for pure states $S=0, \ S_a=S_b$.

We would like to show here an interesting relation between information theory
and hidden variables assumption.
According to hidden variables theory the correlation between the two separated
systems was produced during the interaction time in the past by common hidden
variables. This idea can be represented in the case of two-component
system as [27]:
\begin{eqnarray}
P_{ij} = \sum_\lambda P_{i\lambda} P_{j\lambda} \ ; \ \ 
P_i=\sum_\lambda P_{i\lambda} \ ; \ \ 
P_j=\sum_\lambda P_{j\lambda} 
\end{eqnarray}
where we have assumed that the correlation $P_{ij}$ is produced by common\
hidden variables $\lambda$ for the two separated systems. For simplicity we
assume here a summation over hidden variables $\lambda$, but the present arguments
can be easily generalized to integration over any number of continuous
variables $\lambda$.

Using information theory, we find that hidden variables 
theories lead to a refined distribution [21,22] where the original values 
of $P_i$ and $P_j$ have been resolved into a number of values 
$P_{i\lambda}$ and $P_{j\lambda}$. The resolution of $P_{ij}$ follows, 
however, the constraint that the two separated systems have the common 
parameters $\lambda$.

The refinement of Eq.~(4) can be described as:
\begin{eqnarray}
I^{HV}_{\rm Shann} = \sum_{ij} \sum_\lambda P_{ij\lambda}\ln
\left(\frac{P_{ij\lambda}}{P_iP_jP_\lambda}\right) 
\end{eqnarray}
where
\begin{eqnarray}
\sum_\lambda P_\lambda = 1 \ . \nonumber
\end{eqnarray}
We can now use the log sum inequality [21,25]
\begin{eqnarray}
\sum_\lambda X_\lambda \ln\left(\frac{X_\lambda}{a_\lambda}\right)
\geq
\sum_\lambda X_\lambda \ln \left(\frac{\sum_\lambda X_\lambda}{\sum_\lambda 
a_\lambda}\right) 
\end{eqnarray}
$(X_\lambda  \geq  0, \ a_\lambda  \geq   0$ for all $\lambda$). Using
Eq.~(11) in Eq.~(10) we get:
\begin{eqnarray}
I^{HV}_{\rm Shann} \geq    \sum_{ij}
\left(\sum_\lambda P_{ij\lambda}\right) \ln
\left(\frac{\sum_\lambda P_{ij\lambda}}{\sum_\lambda P_iP_jP_\lambda}\right)
= \sum_{ij} P_{ij}\ln
\left(\frac{P_{ij}}{P_iP_j}\right) \equiv I_{\rm Shann}   
\end{eqnarray}

We have equality in Eq.~(12) if and only if [25]:
\begin{eqnarray}
P_{ij\lambda} = P_\lambda C_{ij}  
\end{eqnarray}
where $C_{ij}$ is independent of $\lambda$. This singular case in which
$\lambda$ does not correlate the $i$ and $j$ subsystems makes redundant the 
hidden variables assumption [Eq.~(9)] and therefore may be discarded. 
Disregarding this singular case we get the result:
\begin{eqnarray}
I^{\rm HV}_{\rm Shann} > I_{\rm Shann} \ ,    
\end{eqnarray}
which means that the refinement by hidden variables should increase
the amount of information included in the Shannon index of correlation. In
particular, if the hidden variables were compatible with other observables
then, for a two-component system in a pure state, measurement of the index of
correlation for the Schmidt representation would provide more information
than half the information contained in the quantum index of correlation.
We find that hidden variable theories lead to mutual 
information between subsystems which is larger than that obtained by QM.
A by-product of refuting the hidden variables theories is reestablishing
the quantum limit of mutual information.\\

\noindent{\bf 3. Empirical State Determination of Entangled Two-Level Systems}

An observable A is represented by an Hermitian operator $A$  
on Hilbert space $H$. The mean value of A, obtained
from an ensemble of systems all prepared identically (described by the 
density matrix $\rho$), 
is given by 
\begin{eqnarray}
\langle A \rangle = {\rm Tr}(\rho A) 
\end{eqnarray}
The problem is to find a set of operators $A$ so that
equations (15) can be solved uniquely for $\rho$ [1,2]. In order to 
determine the density operator of $N$ two-level systems, $2^{2N}-1$ real
numbers are required (because $\rho$ is hermitian and satisfies 
${\rm Tr} \rho = 1$). We need 
therefore $2^{2N}-1$ independent observables by which we can determine 
$\rho$ (and for pure states we have the additional constraint $\rho^2=\rho$).

As is well known [5], the density operator for a two-level system $(N=1)$, 
corresponding to a wavefunction $C_1 | 1 \rangle + C_2 | 2 \rangle$, can be
determined by three measurements: 1) Real part of the ``complex dipole''
$D_1=C^\ast C_2 + C_2^\ast C_1$. 2) Imaginary part of the complex dipole
$D_2=-i(C_1^\ast C_2-C_2^\ast C_1)$. 3) Inversion of population 
$D_3=C_2^\ast C_2 - C_1^\ast C_1$. The condition $\rho^2=\rho$ for pure
states then gives the well known relation $D^2_1 + D_2^2 + D^2_3 = 1$.

For two-spin-$\frac{1}{2}$ systems denoted by $a$ and $b$ we can use the 
Hilbert-Schmidt (HS) representation of the density operator [28]:
\begin{eqnarray}
\rho = \frac{1}{4} \left\{(I)_a \otimes (I)_b +
(\vec{r}\cdot \vec{\sigma})_a \otimes (I)_b + (I)_a \otimes (\vec{s}\cdot
\vec{\sigma})_b + \sum^3_{m,n=1} t_{nm}(\sigma_n)_a
\otimes(\sigma_m)_b)\right\} 
\end{eqnarray}
Here $I$ stands for the unit operator, $\vec{r}, \vec{s}$ belong to
$\RR^3$, $\{\sigma_n\}$ $(n=1,2,3)$ are the standard Pauli matrices. The
coefficients $t_{mn} = Tr(\rho\sigma_n\bigotimes\sigma_m)$ form a real matrix
denoted by $T$. In order to obtain $(r_i)_a$ or $(s_j)_b$ one needs to
perform a measurement on one corresponding arm of the measuring device,
while the parameters $t_{mn}$ involve correlation measurements which are performed
on the two arms. The set of 15 real numbers separates into two 
different classes: 6 real numbers corresponding to $\vec{r}$ and $\vec{s}$ 
describing local properties of the entangled state, and 9 real numbers 
corresponding to the matrix $T$ describing the EPR correlations [28]. If 
we have only the information that our system is composed of two two-level 
subsystems, we need to measure the expectation values of the above 15 
observables in order to determine the quantum state. However, if we have, 
for example, the previous information that our entangled quantum state is 
given by
\begin{eqnarray}
|\psi\rangle = C_1 |1\rangle_a |2\rangle_b + C_2|2\rangle_a|1\rangle_b \ , 
\end{eqnarray}
and only the complex numbers $C_1$ and $C_2$ are not known (but $|C_1|^2+|C_2|^2=1)$,
then the number of real numbers which should be determined by the quantum
measurements is reduced to 3. The assumption made in the use of Eq.~(17)
is equivalent to having the information $r_i$ $(i=1,2,3)=0$, 
$s_j(j=1,2,3)=0$, $t_{ij}(i\neq j)=0$, remaining with only the unknown 
parameters $t_{11}, t_{22},t_{33}$ (where for pure states 
$t_{11}^2+t^2_{22}+t^2_{33}=3$). The description of the 
Bohm singlet state and its relation 
to quantum measurement has played a major role in the 
interpretation of EPR correlations [7]. The wavefunction of this state is a 
special case of Eq.~(17) in which $C_1=\frac{1}{\sqrt{2}}$, 
$C_2=\frac{-1}{\sqrt{2}}$. The HS representation of the density matrix of 
this state is given by the following sum of direct products 
[24]:
\begin{eqnarray}
\rho = \frac{1}{4}(I)_a\otimes(I)_b 
-\frac{1}{4}(\sigma_1)_a\otimes(\sigma_1)_b-\frac{1}{4}(\sigma_2)_a\otimes(\sigma_2)_b-
\frac{1}{4}(\sigma_3)_a\otimes(\sigma_3)_b 
\end{eqnarray}
Here a straightforward calculation of the HS parameters gives the
values $t_{11}=t_{22}=t_{33}=-1$, and all other parameters are equal to zero.
One should take into account that quantum information is included also in the
parameters which are equal to zero. A straightforward calculation for the
density matrix of the Bohm singlet state gives\\
\begin{eqnarray}
\rho=\left(
\begin{array}{rrrr}
0 & 0 & 0 & 0\\[0.25cm]
0 & \frac{1}{2} & -\frac{1}{2} & 0 \\[0.25cm]
0 & -\frac{1}{2} & \frac{1}{2} & 0 \\[0.25cm]
0 & 0 & 0 & 0 \end{array} \right)
\end{eqnarray}
Although all the quantum information of the Bohm singlet state is included
in Eq.~(19), a more direct relation to quantum measurements and physical 
insight 
into the EPR problem is obtained by the HS decomposition given by Eq.~(18).

Let us formulate the EPR problem by following the ``quantum mystery''
description presented by Mermin for three entangled particles [17], which 
holds here also for two entangled particles: ``In the absence of 
connections between the detectors and the source, a particle has no 
information about how the switch of its detector will be set until it 
arrives there''. ``It would seem to be essential for each particle to be 
carrying instructions for how its detector should flash for either of two 
possible switch settings it might find upon arrival''. We would like to 
show here that QM does not introduce any nonlocality problem. According to 
QM all the measurements which can be made on the subsystems $a$ and $b$ of 
the singlet Bohm state are fixed by the ``instructions'' $\vec{r}, \vec{s}$ 
and ${T}$, 15 parameters with the values 
$t_{11}=t_{22}=t_{33}=-1$ and the other 12 parameters equal to 
zero, which were obtained during the production stage of the entangled 
state. Bell's inequalities have been refuted by applying different sets of 
measurements in the different arms of the measuring device [14-17]. 
Although the representation (18) assumes axes of measurements corresponding 
to $\sigma_1,\sigma_2,\sigma_3$ (which might be defined as the $x,y$ and 
$z$ axes) changes of axes of measurements can be obtained by rotation from 
the basis $F$ to another basis $F'$ [28]:
\begin{eqnarray}
(\vec{\sigma}^{F'})_a = 
{\rm O}_1(\vec{\sigma}^F)_a \ ; \ \ 
(\vec{\sigma}^{F'})_b = {\rm O}_2 
(\vec{\sigma}^F)_b 
\end{eqnarray}

The essential point here is that the rotation of axes in system $a$ can be done
{\em independently} of the rotation of axes in system $b$ so that in addition
to the ``instruction'' obtained by the QM interaction each observer
can rotate individually his axes of measurement. As noted in Ref.~23, the 
assumption of vanishing {\em local} commutators pertaining to {\em single}
systems which is assumed in local-realistic hidden variables theories, is
the one that leads to contradiction with QM. A similar conclusion was 
reached by Mermin: ``This is extremely pleasing, for it is just the fact 
that the $x$ and $y$ components of the spin of a {\em single} particle do 
not commute which leads the well educated quantum mechanician to reject 
from the start the inference of instruction sets.'' In conclusion, 
``instruction sets'' which lead to local-realistic theory are in 
contradiction with QM, but ``instruction sets'' obtained by QM which do not 
violate locality but violate realistic models, even for single particles, 
can be obtained in the HS decomposition of the density matrix.

For the singlet spin system one can assume a common rotation for the 
two entangled subsystems and then the density matrix can be expressed in
the new frame $F'$ with the same expression of Eq.~(16) but with $(\vec{\sigma})^F$ replaced
by $(\vec{\sigma})^{F'}$. This kind of symmetry follows from the equality
$t_{11}=t_{22}=t_{33}$. For a triplet state with $M=0$ we get $t_{33}=-1$,
$t_{11}=t_{22}=1$ and Eq.~(16) will change its form even if it follows a common rotation
O$_1 = $O$_2$.
We find that the symmetry properties and their relation to quantum measurements
are described in a simple way by the HS decomposition [28].

For an entangled state of three two-level particles (denoted by 
$a, b, c$), the HS decomposition becomes:
\begin{eqnarray}
8\rho&=&(I)_a\otimes(I)_b\otimes(I)_c +
(\vec{r}\cdot\vec{\sigma})_a\otimes(I)_b\otimes(I)_c +
(I)_a\otimes(\vec{s}\cdot\vec{\sigma})_b\otimes(I)_c + \nonumber \\
&&+(I)_a\otimes(I)_b\otimes(\vec{p}\cdot\vec{\sigma})_c + 
\sum_{mn} t_{mn}(I)_a\otimes (\sigma_m)_b\otimes(\sigma_n)_c \nonumber 
\\&&+\sum_{k\ell} o_{k\ell}(\sigma_k)_a\otimes(I)_b\otimes(\sigma_\ell)_c
+\sum_{ij} p_{ij}(\sigma_i)_a\otimes(\sigma_j)_b\otimes(I)_c\nonumber \\
&&+\sum_{\alpha,\beta,\gamma} R_{\alpha\beta\gamma}(\sigma_\alpha)_a\otimes
(\sigma_\beta)_b\otimes(\sigma_\gamma)_c 
\end{eqnarray}
We find that the three two-level entangled state is described by 63
parameters: 9 for $\vec{r}$, $\vec{s}$ and $\vec{p}$, 27 for $t_{mn}, \ 
o_{k\ell}$ and $p_{ij}$ and 27 for $R_{\alpha\beta\gamma}$. The parameters
$\vec{r}, \ \vec{s}$ and $\vec{p}$ are obtained by measurement on one
arm of the measurement device, $t_{mn}$, $o_{k\ell}$ and $p_{ij}$
are obtained by the measurements on the corresponding two arms of the
measurement device and $R_{\alpha\beta\gamma}$ are obtained by the 
corresponding measurements on the three arms of the measuring device. For example
the parameters for the entangled state
\begin{eqnarray}
|\psi\rangle = \frac{|1\rangle_a |1\rangle_b|1\rangle_c +
|2\rangle_a |2\rangle_b |2\rangle_c}{\sqrt{2}} 
\end{eqnarray}
are given by
$$R_{122}=R_{212}=R_{221}=-1 \ ; \ \ t_{33}=o_{33}=p_{33}=R_{111}=1$$
and all other parameters are equal to zero. Here again the change of axes
of measurement can be made in the systems $a,b$ and $c$ and by independent
rotations for $(\vec{\sigma})_a$, $(\vec{\sigma})_b$ and $(\vec{\sigma})_c$,
respectively.

The Shannon index of correlation can be given for any number of two-level
system by generalizing Eq.~(4):
\begin{eqnarray}
I_{\rm Shann} = \sum_{ijk\ell\cdots} P_{ijk\ell\cdots}\log
\left(\frac{P_{ijk\ell\cdots}}{P_iP_jP_kP_\ell\cdots}\right) 
\end{eqnarray}
and the quantum correlation for such systems is given by generalizing
Eq.~(7)
\begin{eqnarray}
I_c=S_a+S_b+S_c+S_d \cdots - S    
\end{eqnarray}
where $S=0$ for pure quantum states. For the canonical representation of
$n$ particle GHZ system which are maximally correlated one gets [20,28]:
\begin{eqnarray}
S_a=S_b=S_c=S_d = \cdots \ = \log 2, \ \ I_c = n\log 2 
\end{eqnarray}
while for the Shannon index of correlation defined by Eq.~(23) one gets
\begin{eqnarray}
I_{\rm Shann} = (n-1)\log 2 \nonumber \\
\end{eqnarray}
Generally, for these canonical representations there is one bit of 
information less in the Shannon index of correlation relative to that of
the quantum correlation.\\

\noindent{\bf 4. Summary and Conclusions}

In the present study we have followed the ``orthodox approach'', in which
QM describes ensemble averaging, and have related QM results for entangled
two-level systems to information theory. We have described various 
properties of canonical representations which show maximal correlations and 
have discussed for these states the difference between the Shannon index of 
correlation and the quantum index of correlation. It was shown that
hidden variable theories introduce a refinement of the quantum state leading
to a Shannon index of correlation which is larger than that predicted by QM.
By refuting local hidden variable theories one puts a certain limit on the
information that can be transmitted between subsystems of the entangled state.

By using the HS decomposition of entangled two-level systems we have shown
that QM does not violate locality. The parameters of the HS decomposition,
which have been fixed during the interaction time, enable us to predict the
results of any measurement which will be made on the separated subsystems.
The changes of axes of measurement lead to rotations of the spin vectors 
which can be made independently by the observers in different arms of the 
measurement device. Violations of Bell's inequalities or Bell's theorem 
follow from commutation relations for single particles so that the 
assumption of realism is violated and not locality. The HS decomposition 
has been analyzed for some cases including especially the Bohm singlet 
state and GHZ entangled states.\\

\noindent{\bf Acknowledgements}

Y.\ B.-A.\ would like to thank Macquarie University and Professor Barry 
Sanders for their hospitality.
This research has been supported by a Macquarie University Research Grant.
B.\ C.\ S.\ acknowledges the support of the Institute of Theoretical Physics at
Technion. A.\ M.\ was supported by the Fund for 
Promotion of Research at Technion and by the Technion VPR Fund.
\pagebreak

\noindent{\bf References}
\begin{enumerate}
\item W. Band and J.L. Park, ``The Empirical Determination of Quantum 
States'', Foundations of Physics {\bf 1}, 133-144 (1970).
\item J.L. Park and W. Band, ``A general theory of empirical state determination
in quantum physics: Part I'',
Found. Phys. {\bf 1}, 211-226 (1971); W. Band and J.L. Park,  ``A general 
method of empirical state determination in quantum physics: Part II'',
Found. Phys. {\bf 1}, 339-357 (1971).
\item U. Leonhardt, ``Measuring the Quantum State of Light'' (Cambridge 
University Press, Cambridge, 1997).
\item R. Peierls, ``Surprises in Theoretical Physics'' (Princeton University
Press, Princeton, 1979); ``More Surprises in Theoretical Physics'' (Princeton University
Press, Princeton, 1991).
\item L. Allen and J.H. Eberly, ``Optical Resonance and Two Level Atoms'',
(Wiley, New-York, 1975).
\item Y.H. Shih, A.V. Sergienko, M.H. Rubin and C.O. Alley, ``Two-photon entanglement
in type-II parametric down conversion'' Phys.\ Rev.\ A {\bf 50}, 23-28 (1994);
M.\ H.\ Rubin, D.\ N.\ Klyshko, Y.\ H.\ Shih and A.\ V.\ Sergienko, ``Theory of 
two-photon entanglement in type-II optical parametric down-conversion'', 
Phys.\ Rev. A. {\bf 50}, 5122-5133 (1994); Y.H. Shih and A.V. Sergienko, 
``Observation of quantum beating in a simple beam-splitting experiment: 
Two-particle entanglement in spin and space-time'', Phys.\ Rev.\ A{\bf 50}, 
2564-2568 (1994).\item J. Bub, ``Interpreting the Quantum World'' 
(Cambridge University Press, Cambridge, 1997).
\item A.C. Dotson, ``Interpretive principles and the quantum mysteries'', 
Am.\ J.\ Phys.\ {\bf 66}, 970-972 (1998).
\item Y.\ Aharonov, J.\ Anandan and L.\ Vaidman, ``Meaning of the wavefunction'',
Phys.\ Rev.\ A {\bf 47}, 4616-4626 (1993).
\item N.D.H. Das and T. Qureshi, ``Critique of protective measurement'', Phys.\
Rev.\ A{\bf 59}, 2590-2601 (1999).
\item A.\ Barenco, D.\ Deutsch, A.\ Ekert and R.\ Josza, ``Conditional quantum 
dynamics and logic gates'', Phys.\ Rev.\ Lett.\ {\bf 74}, 4083-4086 (1995).
\item V. Scarani, ``Quantum computing'', Am.\ J.\ Phys.\ {\bf 66},
956-960 (1998).
\item C.\ H. Bennet, G.\ Brassard, C.\ Crepeau,
  R.\ Josza, A.\ Peres and W.\ K.\ Wooters,
``Teleporting an unknown quantum state via dual classical and 
Einstein-Podolsky-Rosen channels'', Phys.\ Rev.\ Lett.\ {\bf 70}, 1895-1899 
(1993); D. Bouwmeester, J.-W. Pan, K. Mattle, M. Eibel, H. Weinfurter,
and A. Zeilinger,``Experimental quantum teleportation'', Nature {\bf 390},
575-579 (1997); D. Boschi, S. Branca, F. De-Martini, L. Hardy, and S. 
Popescu, ``Experimental realization of teleporting an unknown pure quantum
state via dual classical and Einstein-Podolsky-Rosen channels'', Phys. Rev.
Lett. {\bf 80}, 1121-1125 (1998); A. Furusawa, J.L. Sorensen, S.L. 
Braunstein, C.A. Fuchs, H.J. Kimble, and E.S. Polzik, ``Unconditional
quantum teleportation'', Science {\bf 282}, 706-709 (1998).
\item A. Aspect, J. Dalibard and E. Roger, ``Experimental test of 
Bell's inequalities using time-varying analyzers'', Phys.\ Rev.\ Lett.\ 
{\bf 49}, 1804-1807(1982).
\item D.\ M.\ Greenberger, M.\ A.\ Horne, A.\ Shimony and A.\ Zeilinger, 
``Bell's theorem without inequalities'', Am.\ J.\ Phys.\ {\bf 58}, 1131-1143 
(1990).\item N.D. Mermin, ``Extreme quantum entanglement in a superposition
of macroscopically distinct states'', Phys.\ Rev.\ Lett.\ {\bf 65},\
1838-1840 (1990).
\item N.D. Mermin, ``Quantum mysteries revisited'', Am.\ J.\ Phys.\ {\bf 58},
731-734 (1990).
\item E. Hagley, X. Maitre, G. Nogues, C. Wunderlich, M. Brune, J.M. Raimond
and S. Haroche, ``Generation of Einstein-Podolsky-Rosen pairs of atoms'',
Phys.\ Rev.\ Lett. {\bf 79}, 1-5 (1997).
\item B.\ C.\ Sanders, ``Entangled coherent states'', Phys.\ Rev.\ A.\ {\bf  
45},6811-6815 (1992); {\bf 46}, 2966;
B.\ C.\ Sanders, K.\ S.\ Lee and M.\ S.\ Kim, ``Optical 
homodyne measurements and entangle coherent states'' Phys.\ Rev.\  A, 
{\bf 52}, 735-741 (1995).
\item S.\ M.\ Barnett and S.\ D.\ J.\ Phoenix, ``Information theory, squeezing and 
quantum correlations'' Phys.\ Rev.\ A {\bf 44}, 535-545 (1991).
\item H. Everett III, ``The theory of universal 
wave function'' in the ``Many-Worlds interpretation of quantum mechanics'', 
Edit. B.S. DeWitt and N. Graham (Princeton University Press, Princeton, 
1973).
\item H.\ Everett III, ``Relative state formulations of quantum mechanics'',
Rev.\ Mod.\ Phys.\ {\bf 29}, 454-462 (1957).
\item W.\ De Baere, A.\ Mann and M.\ Revzen, ``Locality and Bell's theorem'',
Found.\ Physics {\bf 29}, 67-77 (1999).
\item A. Peres, ``Quantum Theory: Concepts and Methods'' (Kluwer, 
Dordrecht, 1995).
\item T.M. Cover and J.A. Thomas, ``Elements of Information Theory'', 
Chapter 2 (Wiley, New York, 1991).
\item R.G. Gallager, ``Information Theory and Reliable Communication'' 
(Wiley,
New-York, 1968).
\item J.S. Bell, ``Speakable and Unspeakable in Quantum Mechanics'' 
(Cambridge University Press, Cambridge, 1987).
\item R. Horodecki and P. Horodecki, ``Perfect correlations in the 
Einstein-Podolsky-Rosen experiments and Bell's inequalities'',
Physics Letters A {\bf 210}, 227-231 (1996).
\end{enumerate}

\end{document}